\documentclass[reprint, twocolumn, showpacs,amsmath,amssymb,letterpaper,aps,prb, superscriptaddress]{revtex4}
\usepackage{graphicx}
\usepackage{times}
\usepackage{textcomp}
\usepackage{amsmath}
\usepackage{amssymb}
\usepackage{array}

\begin{document}

\preprint{}

\title{Probing the superconductivity of PrPt$_4$Ge$_{12}$ through Ce substitution}

\author{K. Huang}
\affiliation{Department of Physics, University of California, San Diego, La Jolla, California 92093, USA}
\affiliation{Center for Advanced Nanoscience, University of California, San Diego, La Jolla, California 92093, USA}
\affiliation{Materials Science and Engineering Program, University of California, San Diego, La Jolla, California 92093, USA}
\author{L. Shu}
\altaffiliation[Present Address: ]{Department of Physics, Fudan University, Shanghai, China, 200433}
\affiliation{Department of Physics, University of California, San Diego, La Jolla, California 92093, USA}
\affiliation{Center for Advanced Nanoscience, University of California, San Diego, La Jolla, California 92093, USA}
\author{I. K. Lum}
\affiliation{Department of Physics, University of California, San Diego, La Jolla, California 92093, USA}
\affiliation{Center for Advanced Nanoscience, University of California, San Diego, La Jolla, California 92093, USA}
\affiliation{Materials Science and Engineering Program, University of California, San Diego, La Jolla, California 92093, USA}
\author{B. D. White}
\affiliation{Department of Physics, University of California, San Diego, La Jolla, California 92093, USA}
\affiliation{Center for Advanced Nanoscience, University of California, San Diego, La Jolla, California 92093, USA}
\author{M. Janoschek}
\altaffiliation[Present Address: ]{MPA-CMMS Los Alamos National Laboratory, Los Alamos, New Mexico 87545, USA}
\affiliation{Department of Physics, University of California, San Diego, La Jolla, California 92093, USA}
\affiliation{Center for Advanced Nanoscience, University of California, San Diego, La Jolla, California 92093, USA}
\author{D. Yazici}
\affiliation{Department of Physics, University of California, San Diego, La Jolla, California 92093, USA}
\affiliation{Center for Advanced Nanoscience, University of California, San Diego, La Jolla, California 92093, USA}
\author{J. J. Hamlin}
\altaffiliation[Present Address: ]{Department of Physics, University of Florida, Gainesville, Florida 32611, USA}
\affiliation{Department of Physics, University of California, San Diego, La Jolla, California 92093, USA}
\affiliation{Center for Advanced Nanoscience, University of California, San Diego, La Jolla, California 92093, USA}
\author{D. A. Zocco}
\altaffiliation[Present Address: ]{Institute of Solid State Physics (IFP), Karlsruhe Institute of Technology, D-76021 Karlsruhe, Germany}
\affiliation{Department of Physics, University of California, San Diego, La Jolla, California 92093, USA}
\affiliation{Center for Advanced Nanoscience, University of California, San Diego, La Jolla, California 92093, USA}
\author{P.-C. Ho}
\affiliation{Department of Physics, California State University Fresno, Fresno, California 93740, USA}
\author{R. E. Baumbach}
\altaffiliation[Present Address: ]{MPA-CMMS Los Alamos National Laboratory, Los Alamos, New Mexico 87545, USA}
\affiliation{Department of Physics, University of California, San Diego, La Jolla, California 92093, USA}
\affiliation{Center for Advanced Nanoscience, University of California, San Diego, La Jolla, California 92093, USA}
\author{M. B. Maple}
\altaffiliation[Corresponding Author: ]{mbmaple@ucsd.edu}
\affiliation{Department of Physics, University of California, San Diego, La Jolla, California 92093, USA}
\affiliation{Center for Advanced Nanoscience, University of California, San Diego, La Jolla, California 92093, USA}
\affiliation{Materials Science and Engineering Program, University of California, San Diego, La Jolla, California 92093, USA}



\date{\today}

\begin{abstract}
We report measurements of electrical resistivity, magnetic susceptibility, specific heat, and thermoelectric power on the system Pr$_{1-x}$Ce$_{x}$Pt$_4$Ge$_{12}$. Superconductivity is suppressed with increasing Ce concentration up to $x = 0.5$, above which there is no evidence for superconductivity down to 1.1 K.  The Sommerfeld coefficient $\gamma$ increases with increasing $x$ from $\sim$48 mJ/mol K$^2$ up to $\sim$120 mJ/mol K$^2$ at $x$ = 0.5, indicating an increase in strength of electronic correlations. The temperature dependence of the specific heat at low temperatures evolves from roughly $T^{3}$ for $x$ = 0 to e$^{-\Delta/T}$ behavior for $x$ = 0.05 and above, suggesting a crossover from a nodal to a nodeless superconducting energy gap or a transition from multiband to single band superconductivity. Fermi-liquid behavior is observed throughout the series in low temperature magnetization, specific heat, and electrical resistivity measurements. 

\end{abstract}



\maketitle



\section{INTRODUCTION}

Filled skutterudite compounds have been the focus of numerous studies due to the wide variety of strongly correlated electron behavior they exhibit, including Kondo lattice behavior, valence fluctuations, metal-insulator transitions, various magnetically ordered states, spin fluctuations, heavy fermion behavior, non-Fermi liquid behavior, conventional BCS-type and unconventional superconductivity,\cite{SALES02, BAUER02, MAPLE02, MACLAUGHLIN02, AOKI03, VOLLMER03, SUDEROW04, MAPLE05, MAPLE07, SATO09, SHU09} as well as being promising candidates for thermoelectric applications.\cite{SALES96} Filled skutterudites have the chemical formula $MT_{4}X_{12}$ where $M$ can be an alkali metal, alkaline earth, or rare-earth/actinide elements, $T$ = Fe, Os, or Ru, and $X$ = Sb, As, or P.\cite{JEITSCHKO77}

One of the most notable filled skutterudite compounds is PrOs$_{4}$Sb$_{12}$, the first Pr-based heavy fermion superconductor ever reported (previously reported heavy fermion superconductors were Ce- or U-based compounds). The compound has an enormous electronic specific heat coefficient $\gamma$ of $\sim$ 500 mJ/mol K$^2$.\cite{BAUER02, MAPLE02} The specific heat jump at the superconducting critical temperature $T_c$ shows an unusual double peak feature as observed in numerous studies.\cite{MAPLE02, VOLLMER03, MEASSON04} Thermal transport measurements on single crystals carried out as a function of magnetic field direction revealed two different superconducting phases. At high fields (A-phase) the energy gap has four or more point nodes while the low field (B-phase) has two point nodes.\cite{IZAWA03} High field measurements probing the normal state properties revealed the existence of a high field ordered phase,\cite{AOKI02, HO02, HO03, TAYAMA03} which was determined by means of neutron diffraction experiments to be an antiferroquadrupolar ordered state.\cite{KOHGI03} 

A new class of filled skutterudites of the form $R$Pt$_4$Ge$_{12}$ has recently been synthesized,\cite{BAUER07, GUMENIUK08} opening up an entirely new direction for filled skutterudite research. Several members of this new class exhibit superconductivity ($R$ = Sr, Ba, Th, La, Pr) where the $R$ = Pr member has one of the highest values of $T_{c}$ at $\sim$7.9 K.\cite{GUMENIUK08} Recent investigations have suggested that PrPt$_{4}$Ge$_{12}$ exhibits a type of strongly-coupled unconventional superconductivity that has point nodes in the energy gap\cite{MAISURADZE09} and breaks time-reversal symmetry.\cite{MAISURADZE10} 

The Pr-based platinum germanide and osmium antimonide filled skutterudites exhibit a number of similarities and certain differences. Both compounds exhibit evidence for time-reversal symmetry breaking from muon-spin relaxation measurements.\cite{AOKI03, MAISURADZE10} Experiments that probe the superconducting energy gap have yielded evidence for both nodal and nodeless energy gaps in both compounds. Transverse muon spin rotation (TF-$\mu$SR) experiments revealed a temperature dependence of the penetration depth $\lambda$ for PrOs$_{4}$Sb$_{12}$ that is consistent with an isotropic energy gap,\cite{MACLAUGHLIN02} while an NQR study reported evidence that PrPt$_{4}$Ge$_{12}$ is a weakly-coupled BCS superconductor.\cite{KANETAKE10} However, for PrOs$_{4}$Sb$_{12}$ scanning tunneling microscopy measurements observed a gap that was open in large regions, discounting the possibility of line nodes,\cite{SUDEROW04} zero-field microwave penetration depth measurements revealed behavior best described with point nodes in the superconducting energy gap,\cite{CHIA03} and small angle neutron scattering experiments reported distortions in the flux-line lattice that were attributed to gap nodes.\cite{HUXLEY04} Regarding PrPt$_{4}$Ge$_{12}$, transverse field $\mu$SR measurements were best fit by gap nodes.\cite{MAISURADZE09} More recent studies also suggest that PrOs$_{4}$Sb$_{12}$\cite{SEYFARTH05, SEYFARTH06,SHU09} and PrPt$_{4}$Ge$_{12}$\cite{CHANDRA12, NAKAMURA12, ZHANG13} are multiband superconductors. While both compounds have a $\Gamma_1$ singlet ground state, the splitting $\Delta_{\textrm{CEF}}$ between the ground and first excited states differs by an order of magnitude. For PrOs$_4$Sb$_{12}$, $\Delta_{\textrm{CEF}}$ $\approx$ 7 K, \cite{MAPLE06, MAPLE07} while for PrPt$_{4}$Ge$_{12}$, $\Delta_{\textrm{CEF}}$ $\approx$ 130 K.\cite{MAISURADZE09, GUMENIUK08, TODA08} Furthermore, while PrOs$_{4}$Sb$_{12}$ is a heavy fermion compound with an electronic specific heat coefficient $\gamma$ $\sim$ 500 mJ/mol K$^2$, the electronic correlations in PrPt$_{4}$Ge$_{12}$ are considerably weaker as reflected in a much smaller value of $\gamma$ $\sim$ 60 mJ/mol K$^2$.\cite{GUMENIUK08}

In an effort to obtain more insight into the unconventional superconductivity of PrPt$_{4}$Ge$_{12}$, we have performed a detailed study of the evolution of the superconducting and normal state properties of PrPt$_{4}$Ge$_{12}$ when Ce ions are substituted into the filler sites for Pr ions. The objective of these experiments was to determine the relation between the superconducting properties and the magnetic state of the Ce ions inferred from the normal state properties of the Pr$_{1-x}$Ce$_{x}$Pt$_4$Ge$_{12}$ system as $x$ is varied. The resultant behavior of the superconducting properties as a function of substituent concentration are correlated with the magnetic state of the substituent ions which can lead to some extraordinary types of behavior. In the limit $T^{\ast} \ll T_{c_o}$, where $T^{\ast}$ is the characteristic temperature (e.g., Kondo or spin fluctuation temperature) and $T_{c_o}$ is the critical transition temperature of the host superconductor, it has been found that the $T_c$ vs $x$ curve can exhibit reentrant behavior wherein superconductivity that occurs below a certain $T_c$ is destroyed below a second lower $T_c$, whereas, in the limit $T^{\ast} \gg T_{c_o}$, the $T_c$ vs $x$ curve has an exponential shape.



In this paper, we report electrical resistivity, magnetization, specific heat, and thermopower measurements on the Pr$_{1-x}$Ce$_{x}$Pt$_4$Ge$_{12}$ system as a function of Ce concentration $x$ for $0 \le x \le 1$.  Fermi liquid behavior was observed throughout the series and a monotonic suppression of $T_c$ with $x$ is observed up to $x = 0.5$, above which there is no evidence for superconductivity down to 1.1 K. Interestingly, specific heat measurements of the superconductiong state suggest either a crossover from a nodal to a nodeless superconducting energy gap or that PrPt$_4$Ge$_{12}$ is a two band superconductor\cite{ZHANG13} and scattering of electrons by substituted Ce ions suppresses the superconductivity associated with one of the bands.


\section{EXPERIMENTAL DETAILS}

Polycrystalline samples of Pr$_{1-x}$Ce$_{x}$Pt$_{4}$Ge$_{12}$ with $x$ = 0, 0.05, 0.06, 0.07, 0.085, 0.1, 0.14, 0.2, 0.25, 0.3, 0.35, 0.4, 0.45, 0.5, 0.625, 0.75, 0.875, and 1 were synthesized by arc melting in an Ar atmosphere  on a water cooled copper hearth using a Zr getter to minimize oxidation.  The starting materials were obtained from Ce rods (Alfa Aesar 3N, ESPI 3N), Pr ingots (Alfa Aesar 99.9\%), Pt sponge (99.9999+\%), and Ge pieces (Alfa Aesar 99.9999+\%).  The elements were weighed in stochiometric ratios and then arc-melted, turned over, and arc-melted again a total of five times to promote chemical homogeneity.  The samples were then annealed in a sealed quartz tube (containing 200 Torr Ar at room temperature) for 336 hours at 800 $^{\circ}$C. To determine whether Ta foil was a necessary component during the annealing process, extra batches of $x$ = 0.06 and 0.07 were annealed while wrapped in Ta foil. Powder x-ray diffraction (XRD) measurements showed no noticable difference from the batches without Ta foil.

Sample quality was characterized through analysis of powder XRD patterns collected by a Bruker D8 x-ray diffractometer using a Cu K$_{\alpha}$ source. Four-wire electrical resistivity measurements were performed from 300 K to $\sim$1.1 K in a pumped $^4$He cryostat and down to 50 mK using a commercial Oxford Kelvinox $^3$He-$^4$He dilution refrigerator.  Magnetization measurements were performed between 300 K and 2 K in a Quantum Design MPMS equipped with a 7 T superconducting magnet.  Specific heat and thermoelectric power measurements were performed down to 1.8 K using a PPMS DynaCool.  The heat capacity measurement employed a standard thermal relaxation technique.  To measure the thermoelectric power, we applied a static temperature gradient of $\Delta T$/$T$ = $2-5\%$, where the temperature $T$ was measured using commercial Cernox 1050 thermometers and a Lakeshore 340 Temperature Controller.  Copper leads were attached to the sample with silver epoxy in a two-wire configuration.  The DC thermoelectric voltage generated by the sample was measured using a Keithley 2182 Nanovoltmeter and was corrected for a background contribution arising from thermal/compositional asymmetry in the wires running from the sample to the external electronics at room temperature.

\section{RESULTS}


\begin{figure}
\begin{center}
    \includegraphics[width=1\columnwidth]{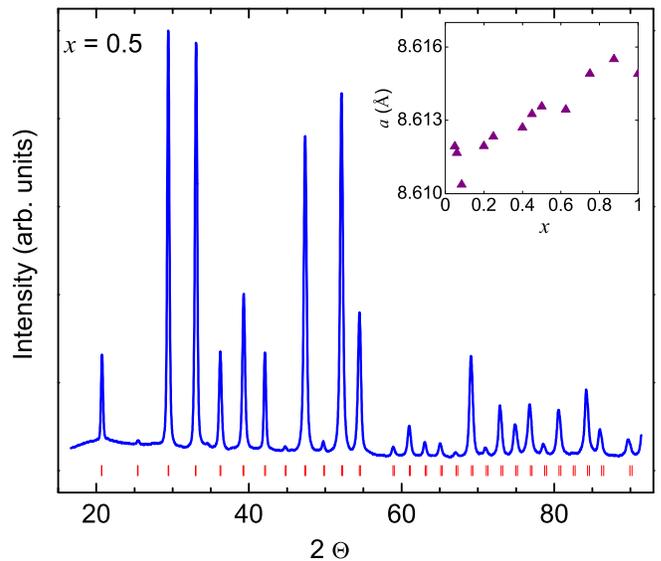}
    \caption{Powder XRD pattern for a representative concentration of Pr$_{1-x}$Ce$_{x}$Pt$_{4}$Ge$_{12}$ ($x$ = 0.5).  Tick marks below the pattern indicate the position of expected Bragg reflections for the refined filled skutterudite crystal structure.  The inset shows a linear increase of lattice parameter $a$ with $x$, which is consistent with Vegard's law.}
    \label{fig:xray}
\end{center}
\end{figure}

Rietveld refinements were performed on powder XRD patterns for each sample using GSAS\cite{LARSON00} and EXPGUI.\cite{TOBY01} The cubic skutterudite crystal structure with space group $Im\bar{3}$ is observed over the entire range of $x$, consistent with expectations.\cite{TODA08, GUMENIUK08} A representative XRD pattern (for the $x = 0.5$ sample) is shown in Fig.~\ref{fig:xray}, with the theoretical peak positions represented as ticks below the XRD pattern.  The agreement between patterns and their refinement with the skutterudite crystal structure was excellent for all samples (fits not shown) with typical reduced $\chi^2$ values around 10. Some samples did show evidence for trace amounts of elemental Ge with impurity concentrations up to 3$\%$, but were otherwise phase pure. Lattice parameters, $a$, obtained for the Pr and Ce parent compounds (inset in Fig.~\ref{fig:xray}) agreed with previous studies.\cite{TODA08, GUMENIUK08} The observed systematic linear increase of $a$ with $x$ is consistent with Vegard's law.

\begin{figure}
\begin{center}
    \includegraphics[width=1\columnwidth]{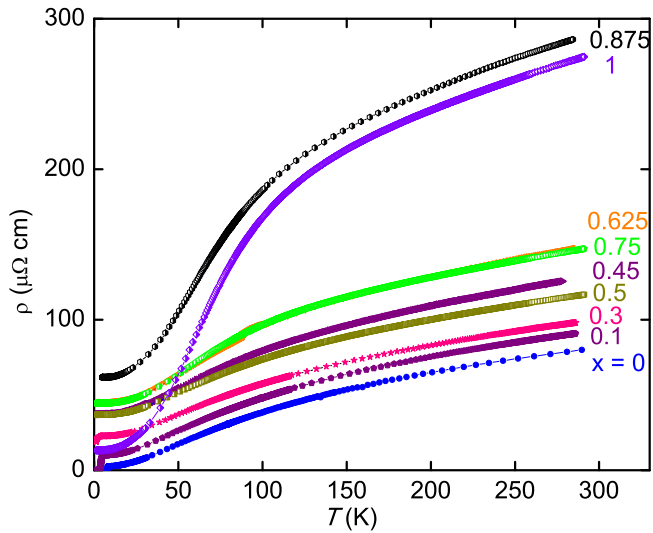}
    \caption{(Color online) Electrical resistivity data for selected Pr$_{1-x}$Ce$_{x}$Pt$_{4}$Ge$_{12}$ samples with Ce concentrations $x$ indicated in figure.}
    \label{fig:rho}
\end{center}
\end{figure}
\begin{figure}
\begin{center}
    \includegraphics[width=1\columnwidth]{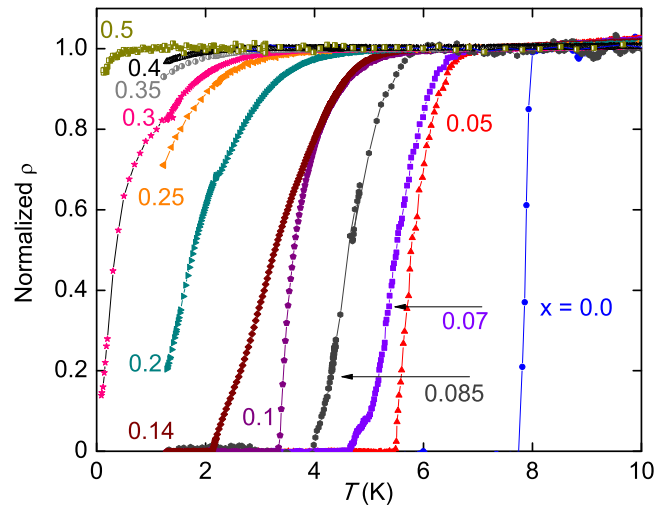}
    \caption{(Color online) Electrical resistivity normalized to the normal state resistivity ($\rho_N$) just above $T_{c}$. $T_c$ decreases with increasing $x$, where the onset of superconductivity is observable up to $x \sim 0.5$.  The transition width broadens with increasing $x$, which may be the result of additional chemical disorder.}
    \label{fig:rho_onset}
\end{center}
\end{figure}

Electrical resistivity, $\rho(T)$, measurements performed in zero applied magnetic field are displayed for representative concentrations in Fig.~\ref{fig:rho}. Data for some concentrations were omitted for the sake of visual clarity. A metallic temperature dependence for $\rho$ was observed for all $x$ as seen in Fig.~\ref{fig:rho}. Significant curvature in $\rho(T)$, which is observed near 80 K for $x$ = 1, is rapdily suppressed as Pr ions are substituted for Ce so that it is no longer observed for $x$ = 0.75 (25$\%$ Pr). By diluting the Ce sub-lattice with Pr ions, which generally do not allow for strong hybridization between localized and itinerant electron states, the scattering contribution related to hybridization between Ce 4$f$ electron states and the conduction band is rapidly destroyed. Gentle curvature in $\rho(T)$ remains for 0 $\le x \le$ 0.75, which may be due to Mott-Jones ''s-d" type scattering. Ce substitution suppresses and broadens the superconducting transition as highlighted in Fig.~\ref{fig:rho_onset}. The onset of superconductivity was observed in samples with concentrations up to $x \sim$ 0.5. The critical temperatures, $T_c$, were identified as the temperature where $\rho$ drops to half its value in the normal state right above the superconducting transition.

    \label{fig:rho_fits}

\begin{figure}
\begin{center}
    \includegraphics[width=1\columnwidth]{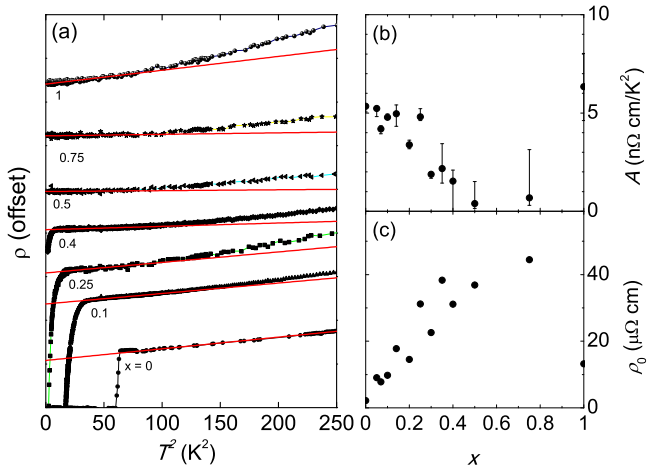}
    \caption{(a) Selected $\rho(T)$ data for Pr$_{1-x}$Ce$_{x}$Pt$_{4}$Ge$_{12}$ plotted as a function of $T^2$, with offsets added for clarity. Power law fits of the form $\rho(T)$ = $\rho_0$ + $AT^2$ were performed up to roughly 250 K$^2$. The solid lines represent best fits to the data. (b) Coefficient $A$ plotted across the entire series. $A$ decreases from 5 n$\Omega$ cm/K$^2$ at $x$ = 0, to a minimum value at $x$ = 0.5, after which it increases. (c) The residual resistivity $\rho_{0}$, extracted from the power law fits, increases with increasing $x$ across the series. $\rho_0$ starts increasing from $\sim$ 1.7 $\mu \Omega$ cm for $x$ = 0 and approaches a maximum of 44 $\mu \Omega$ cm at $x$ = 0.75, after which it drops to 13 $\mu \Omega$ cm at $x$ = 1, due to the great degree of atomic order of the CePt$_4$Ge$_{12}$ end member compound.}
    \label{fig:rho_T^2_phase}
\end{center}
\end{figure}

At low temperatures where phonons are mostly frozen out, $\rho(T)$ is dominated by impurity and electron-electron scattering. As a result, the lattice contribution to the resistivity becomes negligible, reducing the behavior of $\rho(T)$ to the function: $\rho(T) = \rho_0 + AT^n$ where $\rho_0$ is the residual resistivity. For Fermi-liquids, $n$ = 2, while for non-Fermi liquids, $n$ typically lies in the range 0.5 $\le n \le 1.5$ but is usually close to 1.\cite{MAPLE10} For Fermi liquid systems, $A$ is proportional to the square of the electronic specific heat coefficient ($\gamma^2$), which is in turn proportional to the square of the effective mass ($m^*$)$^2$, and inversely proportional to the square of the Fermi temperature ($T_F$)$^2$. For typical metals, $m^*$ is comparable to the mass of the free electron and, therefore, $T_F$ is quite large (e.g., for Cu, $T_F\sim$ 8 x 10$^4$ K).\cite{ASHCROFT76} As a result, the magnitude of $A$ is sufficiently small to make it unfeasible to observe the $T^2$ term experimentally. However, in heavy fermion systems, the strong electronic correlations enhance the effective mass and, in turn, reduce the effective values of $T_F$ so that the $T^2$ term can be readily observed.

Figure~\ref{fig:rho_T^2_phase}(a) displays $\rho$ vs $T^2$, where the selected data sets were offset for clarity. The solid lines are least squares fits to the data with $\rho(T)$ = $\rho_0 + AT^2$ from $T_c$ up to $T^2$ $\sim$250 K$^2$. It should be noted that the samples with $x$ = 0.45, 0.625, and 0.875 were prepared at the same time and exhibited a higher porosity than other samples (30$\%$ less dense). For this reason, they were omitted from the power law analysis. Values for $A$, extracted from the fits, are displayed in Fig.~\ref{fig:rho_T^2_phase}(b) as a function of $x$. $A$ decreases with increasing $x$, from 5.3 n$\Omega$ cm/K$^2$ for $x$ = 0 to a minimum at of $\sim$0.4 n$\Omega$ cm/K$^2$ at $x$ = 0.5, afterwards increasing to 6.3 n$\Omega$ cm/K$^2$ at $x$ = 1. Fig.~\ref{fig:rho_T^2_phase}(c) displays values of $\rho_0$ extracted from the fits as a function of $x$, which increases with $x$ from 1.7 $\mu \Omega$ cm at $x$ = 0 to a maximum of 44 $\mu \Omega$ cm at $x$ = 0.75. A rapid drop occurs in $\rho_0$ at $x$ = 1 down to 13 $\mu \Omega$ cm, which is expected because it is a non-substituted compound and has less disorder, making the overall trend resemble a weighted parabola. The discrepancy of $\rho_0$ for $x$ = 1 with the values reported in Ref.~\onlinecite{NICKLAS12} and Ref.~\onlinecite{GUMENIUK11}  is most likely due to sample quality as the residual resistivity ratio of the previous work was an order of magnitude higher than in the current study.

\begin{figure}
\begin{center}
    \includegraphics[width=1\columnwidth]{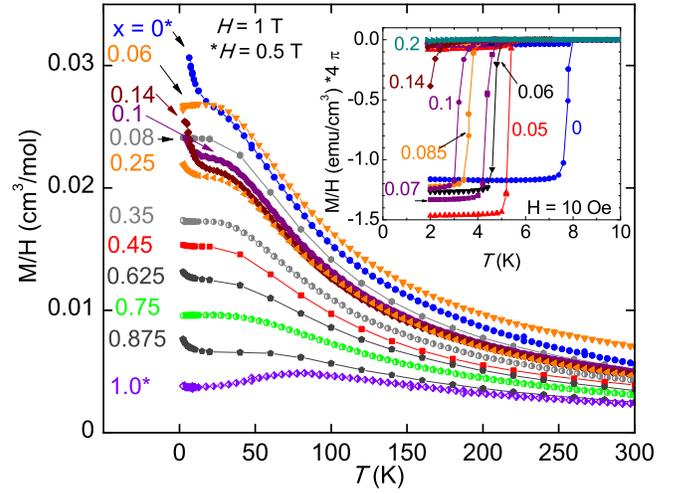}
    \caption{(Color online) $M/H$ as a function of temperature measured in an applied magnetic field of $H$ = 1 T except for the $x$ = 0 and 1 samples which were measured in a $H$ = 0.5 T applied magnetic field.  The inset highlights the Meissner effect in samples with $x \leq 0.2$, for which $M/H$ was measured in an applied field of $H$ = 1 mT.  The superconducting volume fraction appears to be approximately 100$\%$ for $x \leq 0.1$ down to 2 K; however, corrections using the demagnetization factor were not made, which may account for the volume fraction achieving values greater than 1.  $T_c$ decreases with $x$ until $x \sim 0.2$, where no signatures of superconductivity are observed down to 1.8 K.}
    \label{fig:MH}
\end{center}
\end{figure}

Magnetization divided by magnetic field, $M/H$, is displayed as a function of temperature in Fig.~\ref{fig:MH}. The measurements were performed in applied magnetic fields of $H$ = 1 T except for the $x$ = 0 and $x$ = 1 end-member samples which were measured in $H$ = 0.5 T. The magnitude of $M/H$ decreases with Ce concentration throughout the whole series. A broad maximum at $\sim$80 K for $x$ = 1 was also observed. Such a feature could be interpreted as a signature of intermediate valence, consistent with previous studies.\cite{GUMENIUK11} Small upturns in $\chi$($T$) are observed in the low temperature region which appear to be due to small amounts of paramagnetic impurities. Superconducting transition curves for samples with $x \leq 0.2$ are displayed in the inset of Fig.~\ref{fig:MH}, measured in applied magnetic fields of 1 mT to avoid suppressing superconductivity. $T_{c}$ was defined as the temperature in which zero-field cooled (ZFC) and field-cooled (FC) data deviated from one another. The superconducting volume fraction was estimated using the relation $M/H$ $\times$ $d$ = $v$, where $M/H$ has units of emu/mol, $d$ is the density of the compound in units of mol/cm$^3$, and $v$ is the superconducting volume fraction such that $v$ = 1 represents complete flux expulsion by field-induced supercurrents. For samples exhibiting the full transition, the volume fraction achieved values slightly greater than 1, which may be attributed to demagnetization factor effects; nonetheless, the volume fractions are close to 1, indicating bulk superconductivity in this series.

\begin{figure}
\begin{center}
    \includegraphics[width=1\columnwidth]{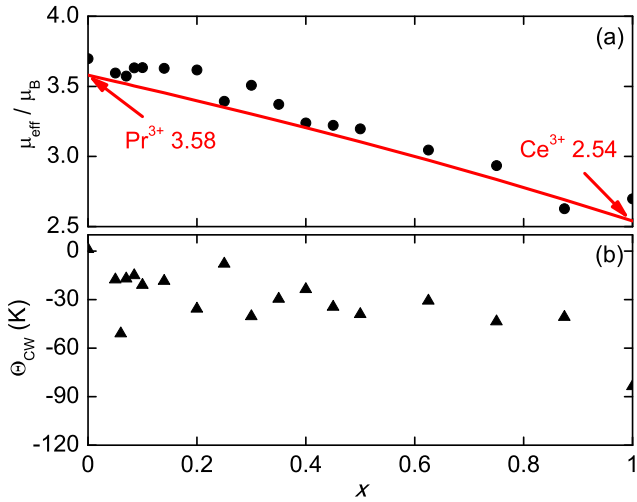}
    \caption{(Color online) Effective magnetic moment ($\mu_{eff}/\mu_B$) and $\Theta_{CW}$ obtained from Curie-Weiss fits to $M/H$ data.  The Pr and Ce parent compounds have $\mu_{eff}$ = 3.69$\mu_{B}$ and  $\mu_{eff}$ = 2.69$\mu_{B}$, respectively, which are close to the expected values for Pr$^{3+}$ and Ce$^{3+}$ free ions (3.58$\mu_{B}$ and 2.54$\mu_{B}$, respectively). The red solid line represents the $\mu_{eff}$ extracted from C-W fits that follows closely the expected values calculated by the relation $\mu_{eff}(x)$ = $\sqrt{(\mu_{Pr3^+})^2(1-x) + (\mu_{Ce3^+})^2(x)}$, which governs the decrease of $\mu_{eff}$ with increasing $x$ when $\Theta_{CW}$ is $x$-independent.}
    \label{fig:ueff_usat}
\end{center}
\end{figure}

The $M/H$ data were fit to a Curie-Weiss law
\begin{equation}
M/H = C_0/(T - \theta_{CW}),
\label{equation:cw}
\end{equation}
in the temperature region 75 - 300 K to determine the Curie-Weiss temperature $\Theta_{CW}$ and average effective magnetic moment $\mu_{eff}$ of the Pr and Ce ions ($\mu_{eff}$ was extracted from the Curie constant $C_0 = \mu_{eff}^2N_{A}/3k_B$, where $N_{A}$ is the number of ions that carry magnetic moments and $k_B$ is Boltzmann's constant). The fits were applied to the $H/M$ data with Eq.~\ref{equation:cw} using linear least squares regression. The resulting best fit values for $\mu_{eff}$ and $\Theta_{CW}$ are displayed in Fig.~\ref{fig:ueff_usat} as a function of $x$. The Pr- and Ce-based end member compounds have effective magnetic moments of 3.69$\mu_{B}$ and 2.69$\mu_{B}$ per lanthanide, respectively, which are consistent with previously reported values.\cite{TODA08, GUMENIUK11} The Curie-Weiss temperature, $\Theta_{CW}$, is roughly independent of $x$, with values $\sim$30 K. Because $\Theta_{CW}$ is nearly constant, the following relation can be used to estimate the expected values of $\mu_{eff}$
\begin{equation}
\mu_{eff}(x) = \sqrt{(\mu_{Pr3^+})^{2}(1-x) + (\mu_{Ce3^+})^{2}(x)},
\label{equation:mu_formula}
\end{equation}
assuming that $\mu_{Pr3^+}$ and $\mu_{Ce3^+}$ have free ion values calculated using Hund's rules (3.58$\mu_{B}$ and 2.54$\mu_{B}$, respectively). The calculated effective magnetic moment $\mu_{eff}$ decreases as a function of $x$ with values (dashed line) that are close to the measured values (solid circles), as can be seen in Fig.~\ref{fig:ueff_usat}(a).

Isothermal magnetization measurements were performed at 2 K as a function of magnetic field (not shown) up to 7 T. Superconductivity was rapidly suppressed; above 1 T, only paramagnetism was observed. Upturns in $M/H$ at low temperature, which may be due to small concentrations of paramagnetic impurities, are most prominent for the samples with $x$ = 0.1 and 0.14. In order to obtain a rough estimate of the concentration of paramagnetic impurities in the samples, the impurities were assumed to be Gd which would be located at the lanthanide sites where Pr and Ce reside. This choice is arbitrary, and we could have chosen another lanthanide such as Ho or Er and Fe, which would occupy the Pt sites. Also, the resulting impurity concentration takes into account impurities on all sites, rare earth or transition metal. The impurity concentration ($N/V$) was determined from Curie law fits to the low temperature upturn using the function $N/V = 3 C_0 k_B/(N_A (\mu_{eff})^2)$, where $\mu_{eff}$ is the effective magnetic moment of Gd (7.94 $\mu_B$), and found to be 1 atomic $\%$ of the lanthanide ions (Pr or Ce). This estimate for the paramagnetic impurity concentration is consistent with values inferred from $M$($H$) isotherms following a procedure described in Ref.~\onlinecite{AONDRAKA95}.



\begin{figure}
\begin{center}
    \includegraphics[width=1\columnwidth]{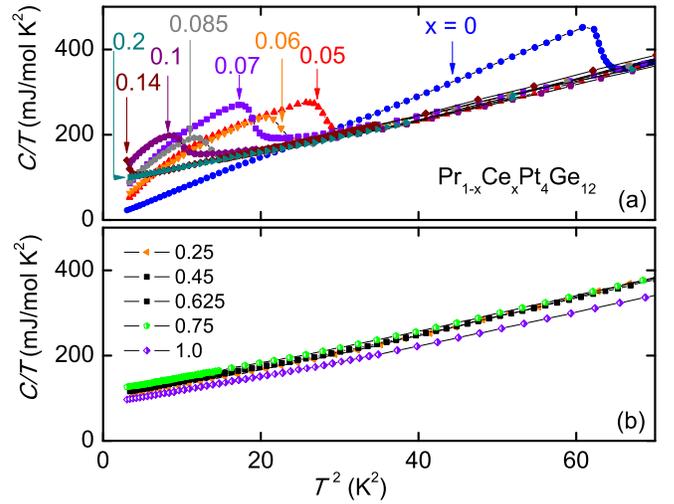}
    \caption{(Color online) Specific heat data displayed as $C$/$T$ vs. $T^2$.  (a) Samples with $x \leq 0.2$ exhibit a jump associated with the transition to the superconducting state.  (b) Samples with $x \geq 0.25$ exhibit no evidence of superconductivity down to 2 K.}
    \label{fig:c_split}
\end{center}
\end{figure}

Specific heat divided by temperature, $C(T)/T$, data are shown in Fig.~\ref{fig:c_split}, where panel (a) displays data for concentrations where superconductivity was observed ($x \le 0.2$), while panel (b) displays data for $x$ $>$ 0.2 for which there was no evidence for superconductivity. The superconducting critical temperature $T_{c}$ was defined as the mid-point of the jump in $C(T)/T$. Consistent with the electrical resistivity and magnetization data, the specific heat data show that superconductivity is suppressed with increasing Ce concentration. The values for $T_c$ extracted from physical properties are addressed in the discussion section.

\begin{figure}
\begin{center}
    \includegraphics[width=1\columnwidth]{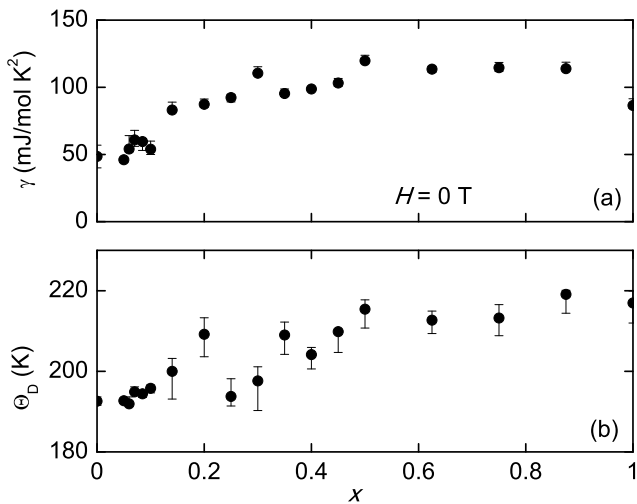}
    \caption{The Sommerfeld coefficient, $\gamma$, and Debye temperature, $\Theta_{D}$, obtained from linear fits of $C(T)$/$T$ = $\gamma$ + $\beta T^2$ to the $C$/$T$ vs $T^2$ data.  The electronic specific heat coefficient $\gamma$ exhibits a moderate enhancement as $x$ increases up to $x = 0.5$, and saturates above this concentration.  The Debye temperature $\Theta_{D}$ increases in magnitude with $x$ throughout the series.}
    \label{fig:c_fits}
\end{center}
\end{figure}

The electronic specific heat coefficient, $\gamma$, and the coefficient of the phonon contribution, $\beta$, were determined from linear fits of $C$/$T$ vs $T^2$ data with the equation $C(T)/T = \gamma + \beta T^2$. The fits were performed from the lowest non-ordered temperature up to as far as linear fits were possible in $C/T$ vs $T^2$. As seen in Fig.~\ref{fig:c_fits}(a), $\gamma$ increases with $x$ from 48 mJ/mol K$^2$ for $x$ = 0 up to roughly 120 mJ/mol K$^2$ for $x$ = 0.5, after which further substitution has a negligible effect on $\gamma$ until $x$ = 1 where it decreases to 86 mJ/mol K$^2$. Although this shows a moderate enhancement of $\gamma$ with increasing $x$, when compared to PrOs$_4$Sb$_{12}$ ($\gamma$$\sim$500 mJ/mol K$^2$), these $\gamma$ values are relatively small. $\gamma$ for the $x$ = 0 sample deviates from values previously reported (87 mJ/mol K$^2$),\cite{GUMENIUK08} yet this may be explained by the differing methods of determining $\gamma$. The authors of Ref.~\onlinecite{GUMENIUK08} suppressed the superconducting region by applying a magnetic field and fitted the temperature region 3-10 K, while in this work the fit was performed on zero field measurements from 7-15 K. The Debye temperature, $\Theta_{D}$, was calculated using the relation: $\Theta_{D} = [1944 \times (n/\beta)]^{1/3}$ K where $n$ = 17,  the number of atoms in the formula unit. $\Theta_{D}$ increases with increasing $x$ for all values of $x$ as seen in Fig.~\ref{fig:c_fits}(b).

\begin{figure}
\begin{center}
    \includegraphics[width=1\columnwidth]{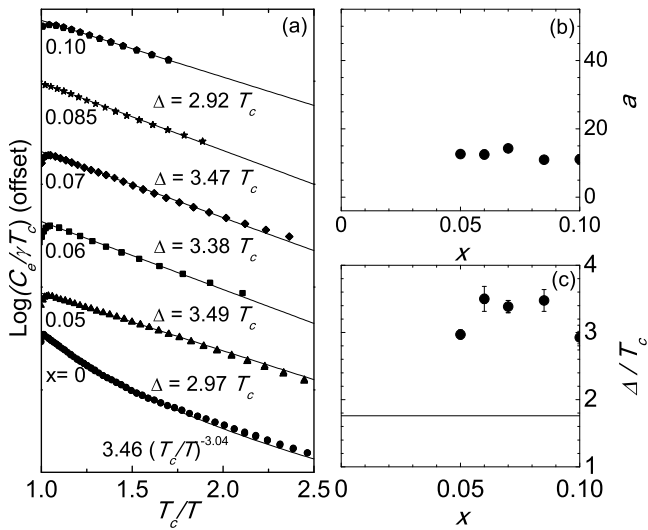}
    \caption{(a) Electronic contribution to specific heat plotted as $\log(C_e/\gamma T_{c})$ vs. $T_{c}/T$. The data were offset for visual clarity. The solid lines represent best fits to the data using 3.46 $T_c/T^{-3.04}$ for $x$ = 0 and $ae^{-\Delta/T}$ for the other concentrations. (b) $a$ vs $x$. (c) $\Delta$/$T_c$ vs $x$. The solid line represents the BCS prediction of $\Delta$/$T_c$ = 1.76.}
    \label{fig:specificHeatTc}
\end{center}
\end{figure}

The electronic contribution to the specific heat, $C_e$, was calculated by subtracting the phonon contribution ($\beta T^3$) from $C$($T$). Figure~\ref{fig:specificHeatTc}(a) shows log($C_{e}$/$\gamma T_c$) vs. $T_c / T$, with data for each concentration offset for visual clarity. The lines in Fig.~\ref{fig:specificHeatTc}(a) represent best fits to the data with a fit range that extends roughly up to $T_c / T$ = 2. As can be seen for samples with $x$ = 0.05 to 0.1, the data are best fit by an exponential of the form $ae^{-(\Delta/T)}$, where $a$ is a fitting parameter and $\Delta$ is the superconducting energy gap. As seen in Fig.~\ref{fig:specificHeatTc}(b-c), the parameters $a$ and $\Delta$/$T_c$ show no clear variation with Ce concentration within the scatter of the data; $a$ ranges in value between 11 and 15, while $\Delta$/$T_c$ ranges from 2.9 to 3.5. However, $\Delta$/$T_c$ is consistently above the BCS prediction of 1.76, indicating that the superconductivity in this series is in the strong-coupling region. There are no data for $a$ and $\Delta$/$T_c$ with $x$ = 0 since it was not possible to fit the $C$($T$) data with the function $ae^{-(\Delta/T)}$. However, as seen in Fig.~\ref{fig:specificHeatTc}(a), fits to the function $a(T_c/T)^n$ where $n\approx$ -3.0 yielded significantly better agreement for the $x$ = 0 sample. The change from power law behavior in the $x$ = 0 sample to exponential behavior in the substituted samples could be explained by a crossover in the superconducting energy gap from a point-node structure for PrPt$_4$Ge$_{12}$ ($\sim T^3$) to a nodeless structure ($e^{-\Delta/T}$) when Ce is introduced.\cite{SIGRIST91, MAISURADZE09} On the other hand, if PrPt$_4$Ge$_{12}$ is a multiband superconductor, another possible explanation is the suppression of one of the superconducting energy bands with the smaller BCS energy gap by external perturbations (in this case, scattering of electrons by substituted Ce ions).\cite{ZHANG13, NAKAMURA12}

    \label{fig:gap}

\begin{figure}
\begin{center}
    \includegraphics[width=1\columnwidth]{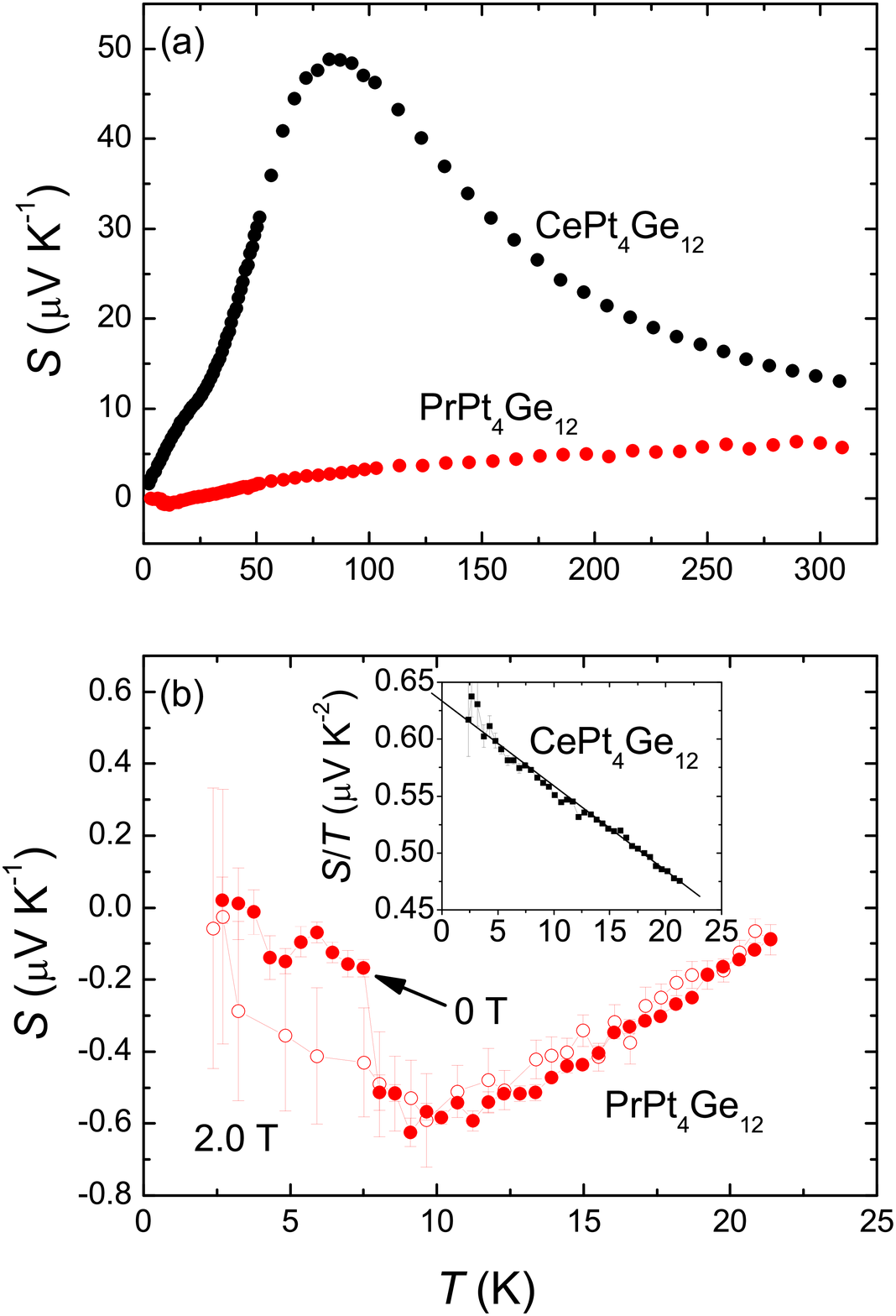}
    \caption{(Color online) (a) Thermoelectric power, $S$, of CePt$_{4}$Ge$_{12}$ and PrPt$_{4}$Ge$_{12}$ as a function of temperature in the absence of applied magnetic field. The compound CePt$_4$Ge$_{12}$ exhibits a broad peak at $\sim$ 80 K as well as a shoulder-like feature at lower temperatures. The presence of the feature suggests that CePt$_{4}$Ge$_{12}$ is on the border of Kondo lattice and intermediate valence. For PrPt$_{4}$Ge$_{12}$, $S$ is roughly 0 $\mu$V/K below $T_c$, as expected for superconductors. (b) A 2 T magnetic field suppresses the superconducting state in PrPt$_{4}$Ge$_{12}$ and leads to a non-zero $S$. The inset displays $S/T$ for CePt$_{4}$Ge$_{12}$, which extrapolates to $\sim 0.63~\mu$V/K$^2$ in the limit $T \rightarrow 0$~K.}
    \label{fig:thermopower}
\end{center}
\end{figure}

Thermopower, $S$, measurements were performed on the $x$ = 0 and 1 compounds and the data are shown in Fig.~\ref{fig:thermopower}. $S$($T$) is roughly 0 $\mu$V/K for the pure Pr compound in the superconducting state below $T_{c}$, as emphasized in Fig.~\ref{fig:thermopower}(b). Applying a 2 T magnetic field suppresses superconductivity and results in a finite value for $S$. $S$ changes sign near 21 K and slowly increases to a broad maximum of $\sim$ 6.3 $\mu$V/K near 300 K. The low magnitudes of $S$($T$) suggests that the electronic density of states for PrPt$_{4}$Ge$_{12}$ may be predominantly flat in the vicinity of the Fermi energy $\epsilon_F$.


For CePt$_{4}$Ge$_{12}$, $S$($T$) goes through a broad peak at approximately 80 K which is consistent with previous reports.\cite{GUMENIUK11} However, the peak observed in this study has a higher magnitude, and the small feature at lower temperatures ($\sim$ 20 K) is more clearly resolved than in previous studies.\cite{GUMENIUK11} The presence of this shoulder-like feature on the large peak indicates that Ce is on the border between Kondo lattice (Ce$^{3+}$) and intermediate valence behaviors within the context of the theory of Zlati\'{c} and Monnier.\cite{ZLATIC05}

\begin{figure}
\begin{center}
    \includegraphics[width=1\columnwidth]{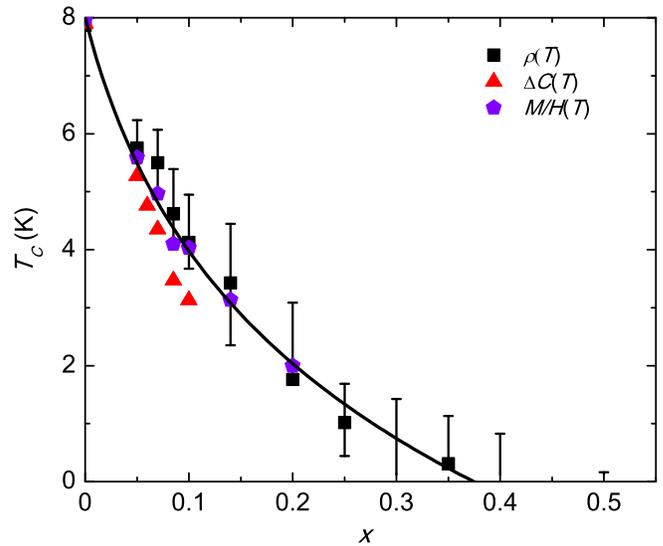}
    \caption{Phase diagram of $T_{c}$ as a function of Ce concentration $x$ obtained from magnetization, specific heat, and electrical resistivity measurements. The error bars for $\rho$ were taken as the 90$\%$ and 10$\%$ drop in resistivity. $T_{c}$ is suppressed by Ce substitution up to $x \sim$ 0.5 with a small positive curvature, above which, evidence for superconductivity is not observed down to $\sim$ 120~mK. In $x$ = 0.3, 0.4, and 0.5 only a portion of the superconducting transition was observed. The solid line is a guide to the eye. }
    \label{fig:phase}
\end{center}
\end{figure}

\section{Discussion}

Summarized in Fig.~\ref{fig:phase} is the evolution of $T_c$ vs $x$ from magnetization, electrical resistivity, and specific heat measurements. All three measurements are consistent with one another in showing that $T_c$ is suppressed with increasing $x$; most evident from electrical resistivity is the positive curvature in the suppression of $T_c$ with $x$. For electrical resistivity and specific heat, the midpoints of the transition were defined as $T_c$, whereas for magnetization the splitting between ZFC and FC was used to determine $T_c$. It would be interesting to compare the evolution of superconductivity due to Ce substitution in PrPt$_4$Ge$_{12}$ to the behavior in PrOs$_4$Sb$_{12}$; an investigation of the Pr$_{1-x}$Ce$_{x}$Os$_{4}$Sb$_{12}$ system is, in fact, underway.

The experiments reported in this paper were undertaken in an effort to probe the nature of the unconventional superconducting state of PrPt$_4$Ge$_{12}$ through the substitution of Ce ions.  In conventional superconductors, it has been shown that the behavior of $T_c$ as a function of the concentration of substituted ions depends sensitively on the magnetic state of the substituent ion.  For Ce ions that are nearly trivalent, as is apparently the case in the Pr$_{1-x}$Ce$_x$Pt$_4$Ge$_{12}$ system, the 4$f$-electron states of the Ce ions are hybridized with conduction electron states and are expected to generate a negative intra-atomic exchange interaction.  The negative exchange interaction should produce a Kondo effect in which the magnetic state of the Ce substituent ion depends on the Kondo temperature $T_K$, wherein the magnetic susceptibility exhibits magnetic behavior (Curie-Weiss behavior) for $T$ $\gg$ $T_K$ and nonmagnetic behavior (Pauli-like behavior) for $T$ $\ll$ $T_K$.  In a conventional superconductor, $T_c$ vs. $x$ changes from a curve with negative curvature that is reentrant in the limit $T_K$ $\ll$ $T_{co}$, where $T_{co}$ is the transition temperature of the superconducting host material (in this case PrPt$_4$Ge$_{12}$), to a curve with positive curvature and nearly exponential shape for $T_K$ $\gg$ $T_{co}$.  The rate of the initial depression of $T_c$ with $x$, -($dT_c$/$dx$)$_{x = 0}$, exhibits a pronounced maximum when $T_{K}$ $\sim$ $T_{co}$.\cite{MULLERHARTMANN71, JARRELL88} The dependence of $T_c$ on Ce concentration for Pr$_{1-x}$Ce$_x$Pt$_4$Ge$_{12}$ shown in Figure~\ref{fig:phase} has positive curvature; the initial linear depression of $T_c$ with $x$ extrapolates to 0 K at $x \sim$ 0.15, whereas the nearly linear region at high $x$ extrapolates to 0 K at $x \sim$ 0.4.  To the extent that Ce ions break superconducting electron pairs in PrPt$_4$Ge$_{12}$, which is apparently an unconventional superconductor with nodes in the energy gap,\cite{MAISURADZE09} the Kondo temperature would appear to be $T_K \sim$ 10-10$^2$ $T_{co}$ $\sim$ 10$^2$-10$^3$ K. Borkowski and Hirschfeld have made a self-consistent theory of Kondo impurities in gapless unconventional superconductors valid in the Fermi liquid regime $T$ $<\sim$ $T_K$.\cite{BORKOWSKI92} However, since the specific heat in the superconducting state quickly develops an exponential shape at low concentrations of Ce, indicative of nodeless superconductivity, we compare our $T_c$ vs. $x$ curves to the theory of Muller-Hartmann and Zittartz, which was developed for Kondo impurities in conventional superconductors with nodeless energy gaps. From this comparison, we estimate that the ratio $T_K$/$T_{co}$ lies in the range 10 - 10$^2$, suggesting that $T_K$ lies in the range 10$^2$ - 10$^3$ K.\cite{MULLERHARTMANN71} For a Kondo temperature with this value, one would expect to observe a minimum in the electrical resistivity that is produced by the sum of the lattice contribution to the resistivity, which decreases with decreasing temperature, and the contribution due to the Kondo effect, which increases with decreasing temperature.  Such a minimum in the electrical resistivity is seen in many dilute alloy systems consisting of a nonmagnetic host metal containing 3$d$ transition metal impurity ions or certain 4$f$ lanthanide (e.g., Ce, Yb) and 5$f$ actinide (e.g., U) impurity ions in which the $f$-electron states are hybridized with conduction electron states.  Examples include the La$_{1-x}$Ce$_x$Al$_2$ system which has a reentrant $T_c$ vs $x$ curve ($T_K \sim$ 0.1 K, $T_{co}$ = 3.3 K) and Th$_{1-x}$U$_x$ which has a nearly exponential $T_c$ vs $x$ curve ($T_K \sim$ 100 K, $T_c$ = 1.4 K).\cite{MAPLE70, MAPLE72, MAPLE76}  However, no resistivity minimum is observed in the Pr$_{1-x}$Ce$_x$Pt$_4$Ge$_{12}$ system, which could be due to the fact that the system is far beyond the single impurity limit. The interactions between the Pr and Ce ions may be sufficiently strong that the normal and superconducting states are determined by the cooperative behavior of the Pr and Ce ions in Pr$_{1-x}$Ce$_x$Pt$_4$Ge$_{12}$.  Thus, it may be more appropriate to think about the gradual evolution of the Pr$_{1-x}$Ce$_x$Pt$_4$Ge$_{12}$ system with increasing x towards CePt$_4$Ge$_{12}$, which is a nonsuperconducting Kondo lattice system with a very large Kondo temperature, which is consistent with the saturation of the magnetic susceptibility as $T$ $\rightarrow$ 0 K.

\begin{table*}
\begin{center}
\caption{Comparison of known Pr$T_4X_{12}$ superconducting skutterudite compounds and their characteristic properties. }
\label{tb:PrScSkutterudites}
\begin{tabular}{l c c c c c c c}
\hline \hline
& $T_c$ & $\mu_{eff}$ & $\gamma$ & $\Delta/k_BT_c$ & $\Theta_{D}$ & $\Delta C/\gamma T_c$ & References\\
& [K] & [$\mu_B$] & [mJ/mol K$^2$] &  & [K] & [mJ/mol K]\\ \hline
PrOs$_4$Sb$_{12}$ & 1.85 & 2.97 & $\sim$500 &  & $\sim$210 & $\sim$1.5 & \onlinecite{BAUER02, MAPLE03, FREDERICK04, MAPLE07}\\
PrRu$_4$Sb$_{12}$ & 1.05 & 3.58 & 59 & 1.53 & 232 & 1.87 & \onlinecite{TAKEDA00, FREDERICK04, MAPLE07}\\
PrRu$_4$As$_{12}$ & 2.5 & 3.52 & 70 & 1.44 & 344 & 1.53 & \onlinecite{NAMIKI07, MAPLE07, SAYLES10}\\
PrPt$_4$Ge$_{12}$ & 7.9 & 3.59 & 87 & 2.35 & 198 & 1.56 & \onlinecite{GUMENIUK08, TODA08, ZHANG13}\\
PrPt$_4$Ge$_{12}$ & 7.9 & 3.69 & 48 &  & 193 & 2.78 & This work\\
\hline \hline
\end{tabular}
\hfill{}
\end{center}
\end{table*}

In view of the unconventional superconductivity exhibited by PrPt$_4$Ge$_{12}$, it is interesting to compare this compound to other Pr based filled skutterudite compounds. Table~\ref{tb:PrScSkutterudites} displays the characteristic parameters of known Pr-based superconducting skutterudites.\cite{MAPLE07} The values of characteristic parameters appear to be very similar when comparing the properties among the majority of these skutterudites. The effective magnetic moment $\mu_{eff}$ remains close to the Hund's rule prediction for the Pr free ion of 3.69$\mu_B$ except for PrOs$_4$Sb$_{12}$, for which $\mu_{eff}$ = 2.97$\mu_B$. The electronic specific heat coefficient $\gamma$ is consistently below 100 mJ/mol K$^2$ with only PrOs$_4$Sb$_{12}$ having a significantly larger value of $\gamma$ $\approx$ 600 mJ/mol K$^2$.\cite{BAUER02, MAPLE03, FREDERICK04, MAPLE07} The Debye temperature $\Theta_D$ exhibits values near 200 K except in PrRu$_4$As$_{12}$ with $\Theta_D$ = 344 K.\cite{NAMIKI07, MAPLE07, SAYLES10} While many of the Pr-based skutterudites have values of $T_c$ between 1-2 K, PrPt$_4$Ge$_{12}$ is unique with $T_c$ = 7.9 K. The ratio $\Delta/k_B T_c$ is also enhanced for PrPt$_4$Ge$_{12}$ relative to the other compounds, exhibiting values of approximately 2.3. It is interesting to observe that many of the properties of these Pr-based superconducting filled skutterudites are very similar, despite the different atomic cages within which the Pr ions reside.

\begin{figure}
\begin{center}
    \includegraphics[width=1\columnwidth]{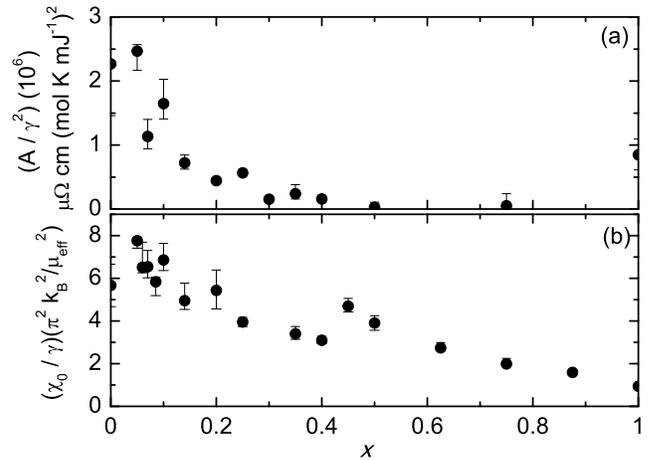}
    \caption{Kadowaki-Woods ratio, $R_{KW}$ = $A/\gamma^2$, and the dimensionless Sommerfeld-Wilson ratio, $R\propto\chi_0/\gamma$, plotted as functions of $x$. The uncertainty in $R_{KW}$ and $R$ were propagated through in calculations from the errors of $A$, $\gamma$, and $\chi_0$ employing standard error analysis.\cite{TAYLOR97} (a) $R_{KW}$ is of order $10^{-6}$ $\mu\Omega$ cm (mol K mJ$^{-1}$)$^2$, which is close to expected values for heavy-fermion systems.\cite{TSUJII05} ($x$ = 0.45, 0.625, and 0.875 were omitted as explained in the text.) (b) $R$ decreases with increasing $x$, suggesting that the Ce parent behaves similarly to a free electron system, while the higher values for small $x$ may be due to magnetic exchange enhancement.}
    \label{fig:KW_SW_ratio}
\end{center}
\end{figure}

Illustrated in Fig.~\ref{fig:KW_SW_ratio}(a) is the Kadowaki-Woods ratio $R_{KW}$ = $A/ \gamma^2$ as a function of $x$, where $A$ is derived from power law fits to $\rho$($T$) with $n$ = 2. The samples of $x$ = 0.45, 0.625, and 0.875 are omitted as there was no $A$ value to calculate $R_{KW}$ with. Figure~\ref{fig:KW_SW_ratio}(b) displays the evolution of the Sommerfeld-Wilson ratio $R$ with $x$, where $R$ = ($\pi^2 k_B^2/(\mu_{eff})^2$) $\chi_0/\gamma$.\cite{GEGENWART05} The error bars of $R_{KW}$ and $R$ in Fig.~\ref{fig:KW_SW_ratio} were propagated through in calculations from the errors bars of $A$, $\gamma$, and $\chi_0$ employing standard error analysis.\cite{TAYLOR97} For many $f$-electron based heavy fermion compounds, $R_{KW}$ = 1.0 x $10^{-5}$$\mu\Omega$ cm(mol K mJ$^{-1})^2$.\cite{KADOWAKI86} However, a number of heavy fermion systems that have Fermi liquid characteristics exhibits values closer to $A/\gamma^2$ = 1.0 x $10^{-6}$, which can be explained by taking into account the degeneracy of the lanthanide ions.\cite{TSUJII05} Fig.~\ref{fig:KW_SW_ratio}(a) displays $R_{KW}$ vs. $x$ where $R_{KW}$ decreases with increasing $x$, from 2.6 x 10$^{-6}$ $\mu\Omega$ cm (mol K mJ$^{-1}$)$^2$ for $x$ = 0 down to 0.05 x 10$^{-6}$ $\mu\Omega$ cm(mol K mJ$^{-1})^2$ for $x$ = 0.5. $R_{KW}$ then increases to 0.85 x 10$^{-6}$ $\mu\Omega$ cm(mol K mJ$^{-1})^2$ for $x$ = 1. For $x$ = 1, $R_{KW}$ is comparable to the value reported by Gumeniuk et al.\cite{GUMENIUK11} However; we were unable to find a reported value for $x$ = 0. $R_{KW}$ stays in the range of 10$^{-6}$, suggesting that Pr$_{1-x}$Ce$_{x}$Pt$_{4}$Ge$_{12}$ behaves similarly to other heavy fermion systems even with an only a modestly enhanced $\gamma$ $\sim$ 110 mJ/mol K$^2$.

For the Sommerfeld-Wilson ratio, $R$, a value of 1 is expected a for free electron gase and a value of 2 for a Kondo system.\cite{FISK87} In Fig.~\ref{fig:KW_SW_ratio}(b), $R$ decreases with increasing $x$ down to roughly 1 at $x$ = 1, suggesting that the Ce parent compound behaves as a free electron system. The higher value for $x$ = 0 may be due to exchange enhancement of the magnetic susceptibility and is consistent with previous literature results for $x$ = 0 ($R\sim$3).\cite{MAISURADZE10, GUMENIUK08}

\section{Concluding Remarks}

A systematic study of the system Pr$_{1-x}$Ce$_{x}$Pt$_{4}$Ge$_{12}$ was performed by electrical resistivity, magnetization, specific heat, and thermopower measurements, where Fermi liquid behavior was observed throughout the series. We find that superconductivity is suppressed with increasing Ce with positive curvature up to $x$ = 0.5, above which no evidence for superconductivity was observed down to 1.1 K. The Sommerfeld coefficient $\gamma$ increases with Ce concentration, from 48 mJ/mol K$^2$ for $x$ = 0 up to a maximum of 120 mJ/mol K$^2$ for $x$ = 0.5, a signature of strengthened electronic correlations. Comparisons of the $C/T$ profile in the superconducting state shows that the $C(T)/T$ data are best described by a $T^3$ dependence for $x$ = 0 and an e$^{-\Delta/T}$ dependence for $x \ge$ 0.05. This may be explained by a crossover from a nodal to nodeless superconducting energy gap or the suppression from multiple to single BCS type superconducting energy bands with increasing Ce concentration.



\begin{acknowledgements}

Sample synthesis and initial screening for superconductivity was supported by the US Air Force Office of Scientific Research under MURI grant No. FA 9550-09-1-0603. Low-temperature measurements were supported by the US National Science Foundation under grant No. DMR 0802478. Measurements of electronic and magnetic properties were supported by the US Department of Energy under grant NO. FG0204-ER46105. The crystal growth equipment was partially supported by the US Department of Energy under grant No. FG02-04-ER46178. M. Janoschek gratefully acknowledges financial support from the Alexander von Humboldt Foundation.

\end{acknowledgements}

\bibliography{PrCePt4Ge12}

\end{document}